# Physical spline for denoising object trajectory data by combining splines, ML feature regression and model knowledge

Jonas Torzewski
*Torc Robotics*

Stuttgart, Germany
jonastorzewski@web.de

GitHub: https://github.com/jonasTorz/physical_spline

***Abstract*— This article presents a method for estimating the dynamic driving states (position, velocity, acceleration and heading) from noisy measurement data. The proposed approach is effective with both complete and partial observations, producing refined trajectory signals with kinematic consistency, ensuring that velocity is the integral of acceleration and position is the integral of velocity. Additionally, the method accounts for the constraint that vehicles can only move in the direction of their orientation. The method is implemented as a configurable python library that also enables trajectory estimation solely based on position data. Regularization is applied to prevent extreme state variations. A key application is enhancing recorded trajectory data for use as reference inputs in machine learning models. At the end, the article presents the results of the method along with a comparison to ground truth data.**

## I. Introduction

A key factor in the success of autonomous driving is the availability of high-quality recorded data to train, develop and validate virtual drivers. To enhance data quality, it is common practice to apply post-processing methods to recorded datasets. Compared to online processing, post-processing benefits from greater computational resources and is not subject to causality constraints. One example of a post-processing method that is exclusively used offline is the forward-backward Kalman filter, which estimates states by sequentially processing data in both forward and backward time directions (for further details, see [1]).

Various types of filters exist, as summarized in Table 1. Some filters are designed solely to smoothen data, while others ensure consistency with physical models.

TABLE I. Different denoising Algorithms

| **Type of model** | **Causal/ Not causal** | |
|---|---|---|
| | ***causal*** | ***Not causal*** |
| Model free, with object list | Low pass filters | Polynom fitting, spline fitting, Centered low pass filters |
| Model based with object list | Kalman filter, Moving horizon estimation | Forward backward Kalmanfilter, physical spline (focus of this paper) |
| Raw sensor data | Sensor fusion, Neural nets, point cloud clustering | Training clustering over multiple time stamp forward backwards |

In this paper we combine model knowledge with spline fitting techniques. Thus, we name the method of this paper physical spline. The result is a parameterized model of trajectories of objects. The parameters of the model are determined by optimization. The optimization is carried out such that one point of the estimated/denoised trajectory depends on both past and future data, i.e. the optimization is not causal. This improves the results compared to a forward backwards iteration with a similar model. The advantage comes at the expense of a more complex derivation and higher computation time. To implement the approach online we would have to implement the method as a moving horizon estimator relying only on past measurement data.

## II. Problem Statement

The goal of the method proposed in this paper is to improve the trajectory data of recorded objects. Each object is reported in a stationary frame with respect to the surface of the earth. The frames are 2D and tangential to the surface of the earth. The origin of each frame is fixed to a specific latitude and longitude value. The object representation is illustrated by (Fig. 1).

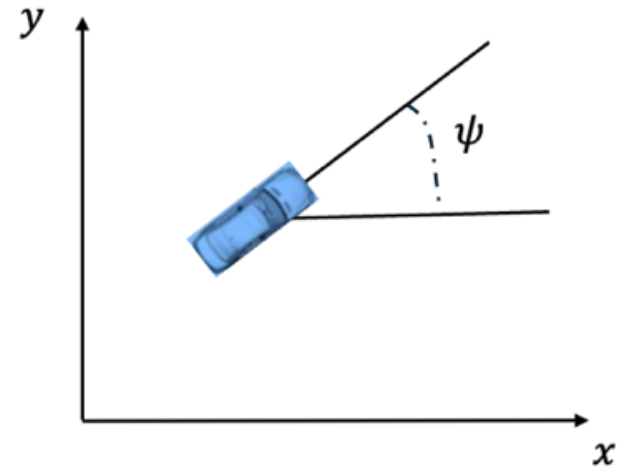

Fig. 1. This figure illustrates how trajectory data is assumed to be given for denoising. Specifically, we assume the data to be in a plane ($x$, $y$). Additionaly, the heading angle $\psi$ of each object is given. For certain use cases derivatives of ($x$, $y$, $\psi$) are also given.

The output of the algorithm we propose consists of differentiable functions $x(t)$, $y(t)$ and $\psi(t)$, where $x(t)$ and $y(t)$ represent the position of the object in the plane and $\psi(t)$ denotes its orientation. The functions shall be determined to be most plausible given some model assumptions and the measured data. The main focus of this article is on determining the functions $x(t)$ and $y(t)$. At the end of the paper the concept will be extended for determining $\psi(t)$.

## III. FUNCTIONAL MODEL OF THE PHYSICAL SPLINE

For estimation purposes, the functions *x(t)* and *y(t)* are modeled as linear combinations of non-linear basis functions

$$x(t) = \sum_j w_{x,j} \cdot f_j(t) \qquad y(t) = \sum_j w_{y,j} \cdot f_j(t)$$
$$\dot{x}(t) = \sum_j w_{x,j} \cdot \dot{f}_j(t) \qquad \dot{y}(t) = \sum_j w_{y,j} \cdot \dot{f}_j(t) \qquad w = \begin{pmatrix} w_{x,0} \\ \vdots \\ w_{x,N} \\ w_{y,0} \\ \vdots \\ w_{y,N} \end{pmatrix}. \tag{1}$$
$$\ddot{x}(t) = \sum_j w_{x,j} \cdot \ddot{f}_j(t) \qquad \ddot{y}(t) = \sum_j w_{y,j} \cdot \ddot{f}_j(t)$$

This approach is also popular in a machine learning context as *"feature based linear regression"* (further details see [2]). The functions will be chosen in such a way that the first three parameters represent the initial state ( $w_{x,0} = x(0), w_{x,1} = \dot{x}(0), w_{x,2} = \ddot{x}(0)$ ). All following parameters ( $w_{x,3}, w_{x,4}, \ldots$ ) represent the acceleration after one resolution step. The parameters in the weight vector have the following representation and are defined on a underlying time vector $t_{grid}$

$$w_x = (x_0 \quad \dot{x}_0 \quad \ddot{x}_0 \quad \ddot{x}_{\Delta t} \quad \ddot{x}_{2\Delta t} \quad \ldots) \tag{2}$$
$$t_{grid} = [t_0, t_1, \ldots, t_j, \ldots, t_N] = [0, \Delta t, 2\Delta t, \ldots].$$

In between the weight centers the acceleration is linear interpolated. Stated otherwise, we choose the basis functions for the second derivatives as first order B-Spline functions [3]. By integrating these basis functions twice we obtain corresponding basis functions for *x(t)* and *y(t)*

$$\ddot{f}_j = \begin{cases} 0 & \text{for } t \le t_{j-1} \\ \dfrac{t - t_{j-1}}{t_j - t_{j-1}} & \text{for } t_{j-1} < t \le t_j \\ -\dfrac{t - t_j}{t_{j+1} - t_j} + 1 & \text{for } t_j < t \le t_{j+1} \\ 0 & \text{for } t > t_{j+1} \end{cases} \tag{3}$$

$$\dot{f}_j = \begin{cases} 0 & \text{for } t \le t_{j-1} \\ \dfrac{\frac{1}{2} \cdot (t - t_{j-1})^2}{t_j - t_{j-1}} & \text{for } t_{j-1} < t \le t_j \\ -\dfrac{\frac{1}{2}(t - t_j)^2}{t_{j+1} - t_j} + (t - t_j) + \frac{1}{2} \cdot (t_j - t_{j-1}) & \text{for } t_j < t \le t_{j+1} \\ \frac{1}{2}(t_{j+1} - t_{j-1}) & \text{for } t > t_{j+1} \end{cases}$$

$$f_j = \begin{cases} 0 & \text{for } t \le t_{j-1} \\ \dfrac{1}{6} \cdot \dfrac{(t - t_{j-1})^3}{t_j - t_{j-1}} & \text{for } t_{j-1} < t \le t_j \\ -\dfrac{\frac{1}{6}(t - t_j)^3}{t_{j+1} - t_j} + \frac{1}{2}(t - t_j)^2 + \frac{1}{2} \cdot (t_j - t_{j-1}) \cdot (t - t_j) + \frac{1}{6} \cdot (t_j - t_{j-1})^2 & \text{for } t_j < t \le t_{j+1} \\ \frac{1}{2}(t_{j+1} - t_{j-1}) \cdot (t - t_{j+1}) + \frac{1}{3}(t_{j+1} - t_j)^2 + \frac{1}{6}(t_j - t_{j-1})^2 + \frac{1}{2}(t_j - t_{j-1})(t_{j+1} - t_j) & \end{cases}.$$

The constants resulting from integration are used to fulfil the condition of the first two weights being the initial position and initial velocity.

$$f_0 = 1 \tag{4}$$
$$f_1 = t - t_0$$

The function $f_0$ has a one in position and a zero in the derivatives. The function $f_1$ has a zero in position and acceleration at $t_0$ and a one in the first derivative. The function $f_2$ denotes the start of the acceleration weights for the initial acceleration.

In (Fig. 2) we can see the plot of the basis functions (except for the first two). We can see that the second derivative, only depends on the corresponding weight of the basis function at its center.

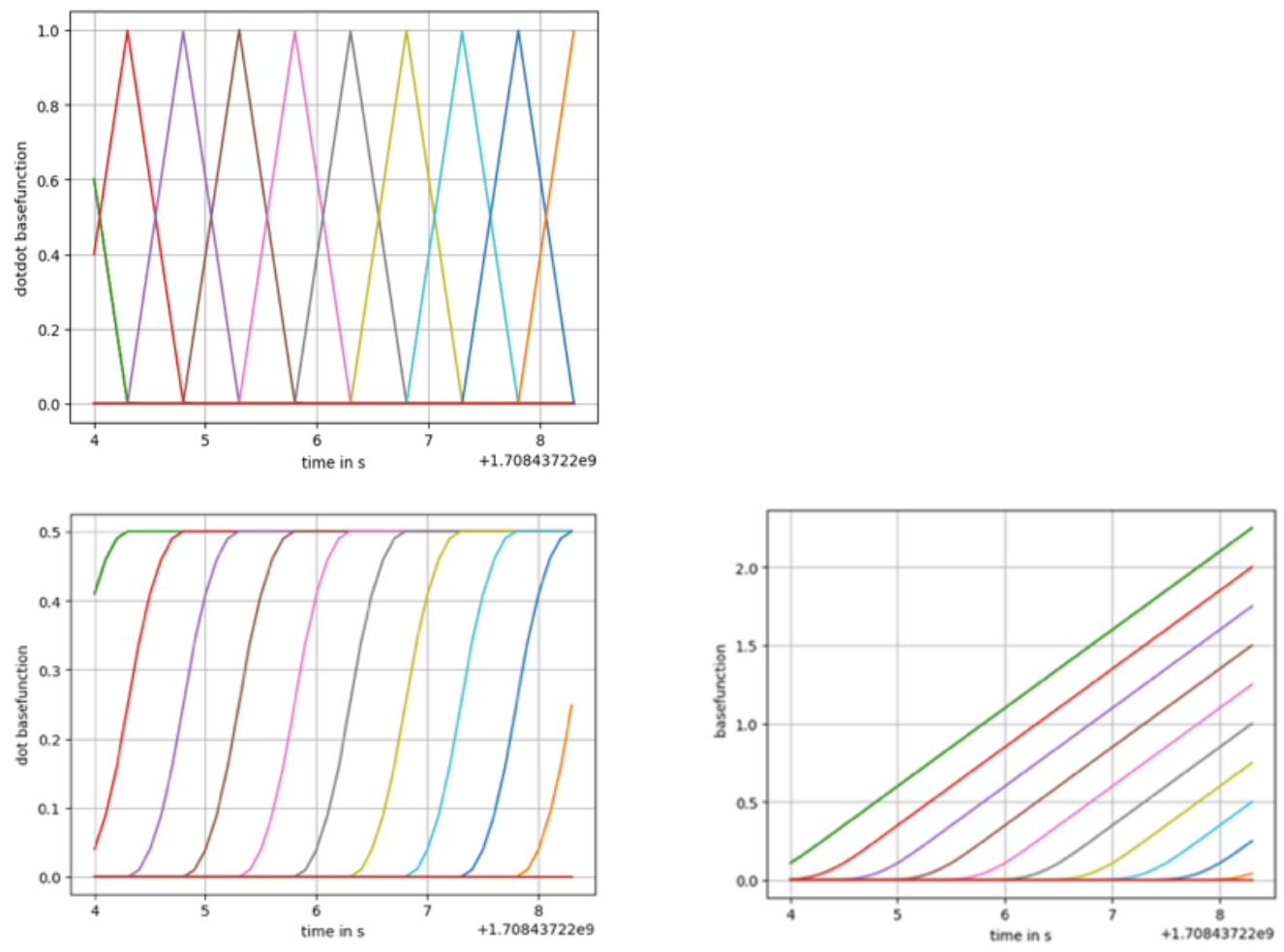

Fig. 2. Basis functions with the first and second derivative. The basis functions look similar to a ReLu, the derivatives to a sigmoid, the second derivative to a radial basis function.

A visual way to think about physical knowledge integrated into the basis functions in (Fig. 2) is that a weight change of one parameter only influences the acceleration function locally through the radial basis function, i.e. the long-term future of the acceleration is not affected by changes in earlier weights. For the velocity a weight change influences the long-term future of the track like a constant offset because the sigmoid like function behaves like a constant for larger times. The position changes linearly by a weight change of an earlier weight. The intuitive picture behind that is that without any action an object in motion stays in motion, the position changes linear, the velocity stays constant, and an acceleration requires constant effort of weights unequal to zero to not go back to zero. If we extend the physical spline by just appending zeros to the parameter vector at the end the vehicle continues to travel at constant speed. This also means that position, velocity and acceleration values at time are not affected by changing parameters for a later point in time (causality). With these concepts in mind more complex models with physics could be incorporated as an extension. Also, an iterative approach to determine the next parameters would be possible without the need to redetermine the parameters of the past.

However in the following we determine the parameter vector with an optimization problem that gets solved once after the whole track is already recorded. In the following we work with the representation from (1) to formulate the Conditions and determine the weight vector.

## IV. Optimization for Parameter Determination

To determine the parameter vector $w$ for a given set of measurements the problem is formulated as a quadratic optimization problem (further details see [4]). The following equations

$$q(w) = \sum q_j(w) \qquad \nabla q(w) = \sum \nabla q_j(w) = 0 \tag{6}$$

$q$: over all cost function
$q_j$: subcost that was added to the overall cost
$w$: parameter vector

show the structure of the cost function and the gradient that needs to be set to zero in order to solve it.

Since the optimization problem to determine $w$ is assumed to be quadratic the gradient of $q(w)$ can be written as Matrix times weight vector plus a vector

$$\nabla q(w) = \sum Q_j \cdot w - b_j = 0 \qquad \underbrace{\sum Q_j}_{Q} \cdot w = \underbrace{\sum b_j}_{b}. \tag{7}$$

$q$: overall cost function
$Q_j$: Matrix of the cost fcn j
$b_j$: vector of the cost fcn j
$w$: parameter vector

Based on this structure different matrices $Q_j$ and vectors $b_j$ are determined and added to the overall matrix $Q$ and vector $b$. Not all of the following elements that contribute to $Q$ and $b$ are needed at the same time. The idea is that we can add multiple of them to the optimization problem and they contribute via their corresponding Matrix $Q_j$ and vector $b_j$.

The next steps are to derive the following costs and add them to the optimization. Specifically, the cost is decomposed as

$$\begin{aligned} q(w) &= q_{position}(w) \\ &+ q_{velocity}(w) \\ &+ q_{heading}(w) \\ &+ q_{lateral,position}(w) \\ &+ q_{longitudinal,position}(w) \\ &+ q_{apriori,regularisation}(w). \end{aligned} \tag{8}$$

### A. L2 Norm to fit positions, velocities and accelerations

In order to consider position, velocities and acceleration measurements, we add up the squared differences between measurement values and the predicted values of the model. Each measurement can be weighted individually by a factor $c_i$

$$\begin{aligned} q_{pos}(w) &= \frac{1}{2}\sum_i^L c_i \cdot (f(t_i) - f_i)^2 = \frac{1}{2}\sum_i^L c_i \left( \left( \sum_j^N f_j(t_i) \cdot w_j \right) - f_i \right)^2 \\ &= \frac{1}{2}\sum_i^L c_i \left( \begin{pmatrix} f_0(t_i) & \dots & f_N(t_i) \end{pmatrix} \cdot \begin{pmatrix} w_0 \\ \vdots \\ w_N \end{pmatrix} - f_i \right)^2 \\ \nabla q_{pos}(w) &= \sum_i^L c_i \left( \begin{pmatrix} f_0(t_i) & \dots & f_N(t_i) \end{pmatrix} \cdot \begin{pmatrix} w_0 \\ \vdots \\ w_N \end{pmatrix} - f_i \right) \cdot \begin{pmatrix} f_0(t_i) \\ \vdots \\ f_N(t_i) \end{pmatrix} \\ &= \underbrace{\sum_i^L c_i \begin{pmatrix} f_0(t_i) \\ \vdots \\ f_N(t_i) \end{pmatrix} \begin{pmatrix} f_0(t_i) & \dots & f_N(t_i) \end{pmatrix}}_{Q_{pos}} \cdot \begin{pmatrix} w_0 \\ \vdots \\ w_N \end{pmatrix} - \underbrace{\sum_i^L c_i \cdot f_i \begin{pmatrix} f_0(t_i) \\ \vdots \\ f_N(t_i) \end{pmatrix}}_{b_{pos}}. \end{aligned} \tag{9}$$

With the same approach the matrices and vectors for velocities and acceleration follow to

$$\underbrace{\sum_i^L c_i \begin{pmatrix} \dot{f}_0(t_i) \\ \vdots \\ \dot{f}_N(t_i) \end{pmatrix} \begin{pmatrix} \dot{f}_0(t_i) & \dots & \dot{f}_N(t_i) \end{pmatrix}}_{Q_{vel}} \cdot \begin{pmatrix} w_0 \\ \vdots \\ w_N \end{pmatrix} - \underbrace{\sum_i^L c_i \cdot \dot{f}_i \begin{pmatrix} \dot{f}_0(t_i) \\ \vdots \\ \dot{f}_N(t_i) \end{pmatrix}}_{b_{vel}}. \tag{10}$$

And for the acceleration to

$$\underbrace{\sum_i^L c_i \begin{pmatrix} \ddot{f}_0(t_i) \\ \vdots \\ \ddot{f}_N(t_i) \end{pmatrix} \begin{pmatrix} \ddot{f}_0(t_i) & \dots & \ddot{f}_N(t_i) \end{pmatrix}}_{Q_{acc}} \cdot \begin{pmatrix} w_0 \\ \vdots \\ w_N \end{pmatrix} - \underbrace{\sum_i^L c_i \cdot \ddot{f}_i \begin{pmatrix} \ddot{f}_0(t_i) \\ \vdots \\ \ddot{f}_N(t_i) \end{pmatrix}}_{b_{acc}}. \tag{11}$$

In the following the weight of an individual Point will not be written for each equation.

### B. Splitting into longitudinal and lateral error

Often the required longitudinal and lateral accuracy is different. Typically, the lateral accuracy must be higher, because that decides if an object is on the lane or not. Also, the measurement quality can be better for lateral positions, because the lane is known and printed on the road and can be used as reference for lateral positions.

Thus, we want to have the option to weigh lateral and longitudinal errors individually. To keep the structure of the optimization problem the same we use a known heading at each measurement point and rotate into the object coordinate system. The resulting error functions are

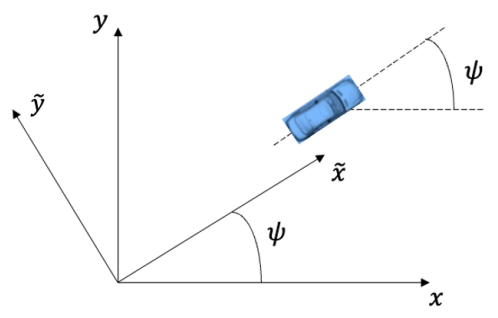

$$\begin{pmatrix} \tilde{x} \\ \tilde{y} \end{pmatrix} = \begin{pmatrix} \cos(\psi) & \sin(\psi) \\ -\sin(\psi) & \cos(\psi) \end{pmatrix} \cdot \begin{pmatrix} x \\ y \end{pmatrix} \tag{12}$$

$$\begin{aligned} e_{lon} &= \cos(\psi_i) \cdot x(t_i) + \sin(\psi_i) \cdot y(t_i) - (\cos(\psi_i) \cdot x_i - \sin(\psi_i) \cdot y_i) \\ e_{lat} &= -\sin(\psi_i) \cdot x(t_i) + \cos(\psi_i) \cdot y(t_i) - (-\sin(\psi_i) \cdot x_i + \cos(\psi_i) \cdot y_i). \end{aligned}$$

The resulting cost functions are

$$\begin{aligned} \frac{1}{2}\sum_i e_{lon,i}^2 &= \frac{1}{2}\sum_i \Big( \cos(\psi_i) \cdot \sum_k w_{k,x} \cdot f_{k,x}(t_i) + \sin(\psi_i) \cdot \sum_k w_{k,y} \cdot f_{k,y}(t_i) \\ &\quad - (\cos(\psi_i) \cdot x_i + \sin(\psi_i) \cdot y_i) \Big) \\ \frac{1}{2}\sum_i e_{lat,i}^2 &= \frac{1}{2}\sum_i \Big( -\sin(\psi_i) \cdot \sum_k w_{k,x} \cdot f_{k,x}(t_i) + \cos(\psi_i) \cdot \sum_k w_{k,y} \cdot f_{k,y}(t_i) \\ &\quad - (-\sin(\psi_i) \cdot x_i - \cos(\psi_i) \cdot y_i) \Big). \end{aligned} \tag{13}$$

Doing the same procedure as above the resulting matrices $Q_j$ and $b_j$ are

Longitudinal:

$$q_i = \begin{pmatrix} \cos(\psi_i) \cdot f_{1,x} \\ \vdots \\ \cos(\psi_i) \cdot f_{n,x} \\ \sin(\psi_i) \cdot f_{1,y} \\ \vdots \\ \sin(\psi_i) \cdot f_{n,y} \end{pmatrix} \qquad \begin{aligned} Q_i &= c_i \cdot q_i \cdot q_i^T \\ b_i &= c_i \cdot (\cos(\psi_i) \cdot x_i + \sin(\psi_i) \cdot y_i) \cdot q_i \\ Q_i &= c_i \cdot q_i \cdot q_i^T \\ b_i &= c_i \cdot (-\sin(\psi_i) \cdot x_i + \cos(\psi_i) \cdot y_i) \cdot q_i \end{aligned} \tag{14}$$

Lateral:

$$q_i = \begin{pmatrix} -\sin(\psi_i) \cdot f_{1,x} \\ \vdots \\ -\sin(\psi_i) \cdot f_{n,x} \\ \cos(\psi_i) \cdot f_{1,y} \\ \vdots \\ \cos(\psi_i) \cdot f_{n,y} \end{pmatrix} \qquad Q_j = \sum Q_i \quad b_j = \sum b_i.$$

The same could be done for the derivatives by using the basis function derivatives. In (Fig. 3) we can see the effect of a longitudinal position jump, if we weigh the longitudinal error low. Also as a human driver we can guess the lateral position of a car that is further away better than its distance, because we can compare it to lane lines and other structures.

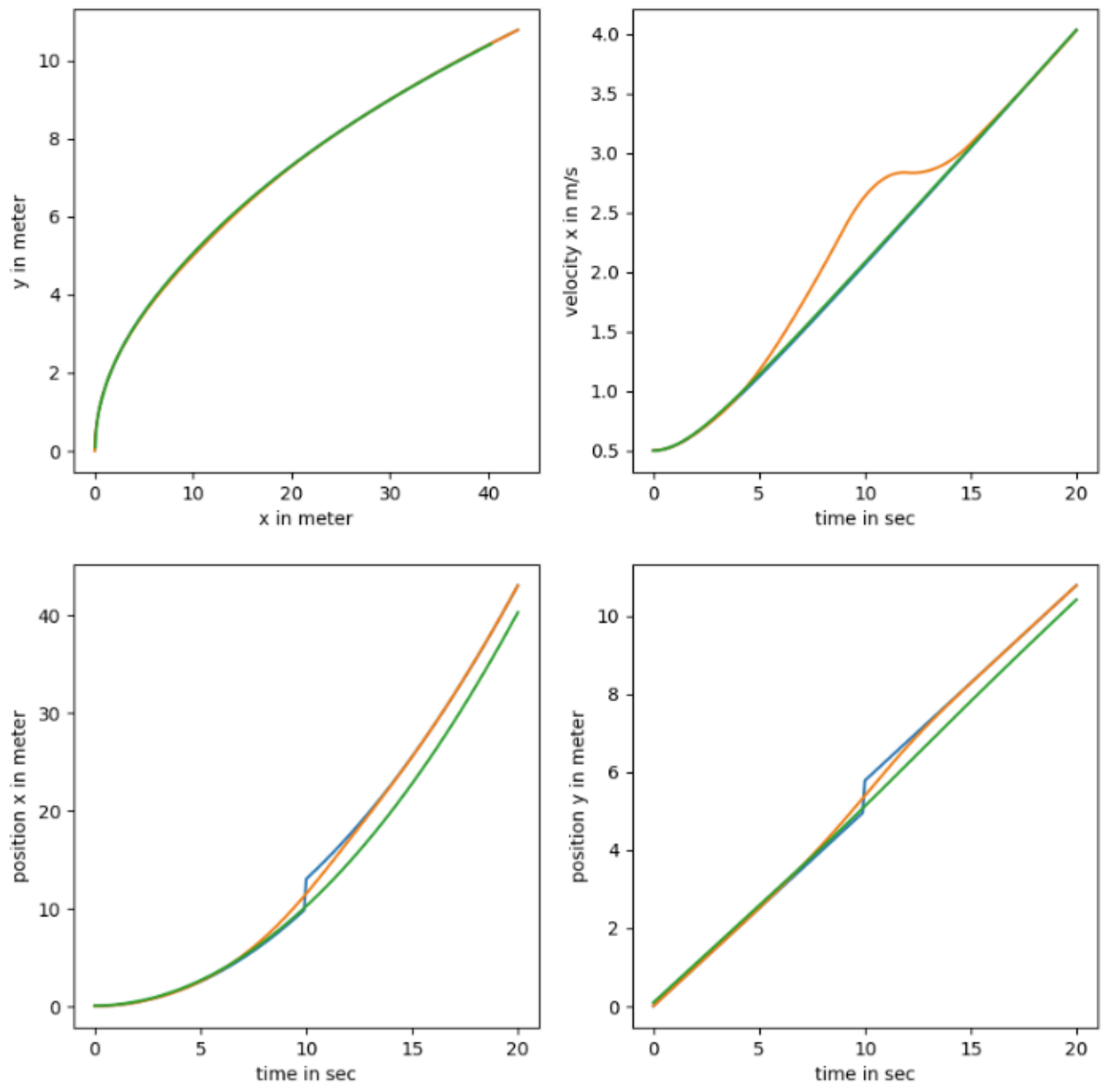

Fig. 3. *Top left birds eye plot, top right velocity over time, bottom left x-position over time, bottom right y-position over time. In orange we added points normally. In green the longitudinal error gets a 1000 times lower weight than the lateral error. Together with the regularization (see below) that results in only a small reaction to the longitudinal jump of the blue input data. We can also see the kinematic consistency in the velocity plot. The orange model has to accelerate in order to move towards the jumped position.*

## C. *Considering heading information*

A vehicle has typically an orientation and can only move into this heading direction (Some deviations are possible especially in drifting situations). In order to get better results we make use of this information and prefer velocities that point into heading direction.

Above it was mentioned that the optimization problem would become nonlinear if we estimate the heading at the same time as the cartesian states. However, it is still possible to consider heading measurements that come from any source as long as we keep the nonlinearity static. To do that we only consider heading measurements for the x and y estimation, but do not estimate the heading itself at the same time. We start with the following equation

$$\frac{\dot{y}}{\dot{x}} = \tan(\psi) \rightarrow 0 = (\dot{x} \cdot \tan(\psi) - \dot{y}) \quad \text{alternativly}: 0 = (-\dot{x} + \dot{y} \cdot \cot(\psi)). \tag{15}$$

The resulting error function looks like (16) . For each point either the tan or cot is chosen depending on which function provides the lower absolute value to avoid the singularities of the functions. By plugin in the formulars of $\dot{x}(t)$ and $\dot{y}(t)$ from (1) and summing over all measurements we get

$$e_{heading} = \frac{1}{2}\sum_j^N c_j\left(\dot{x} \cdot \tan(\psi_j) - \dot{y}\right)^2 \tag{16}$$

$$e = \frac{1}{2}\sum_j^N c_j \left( \tan(\psi_j) \cdot \underbrace{\sum w_{x,i} \cdot \dot{f}_{x,i}(t_j)}_{\dot{x}(t_j)} - \underbrace{\sum w_{y,i} \cdot \dot{f}_{y,i}(t_j)}_{\dot{y}(t_j)} \right)^2 .$$

Using the same procedure as above we get the following matrices

$$q_i = \begin{pmatrix} \tan(\psi_j)\, \dot{f}_{x,1} \\ \vdots \\ \tan(\psi_j)\, \dot{f}_{x,n} \\ -\dot{f}_{y,1} \\ \vdots \\ -\dot{f}_{y,n} \end{pmatrix} \; or \;\; q_i = \begin{pmatrix} -\dot{f}_{x,1} \\ \vdots \\ -\dot{f}_{x,n} \\ \cot(\psi_j)\; \dot{f}_{y,1} \\ \vdots \\ \cot(\psi_j)\; \dot{f}_{y,n} \end{pmatrix} \; Q_i = c_i \cdot q_i \cdot q_i^T \quad b_i = 0. \tag{17}$$

In (Fig. 4) we can see position input data from a circle that was overlayed with a sinus wave. By adding the correct heading we can reconstruct the circle itself. By knowing the road heading we can correct movements that do not follow the road geometry

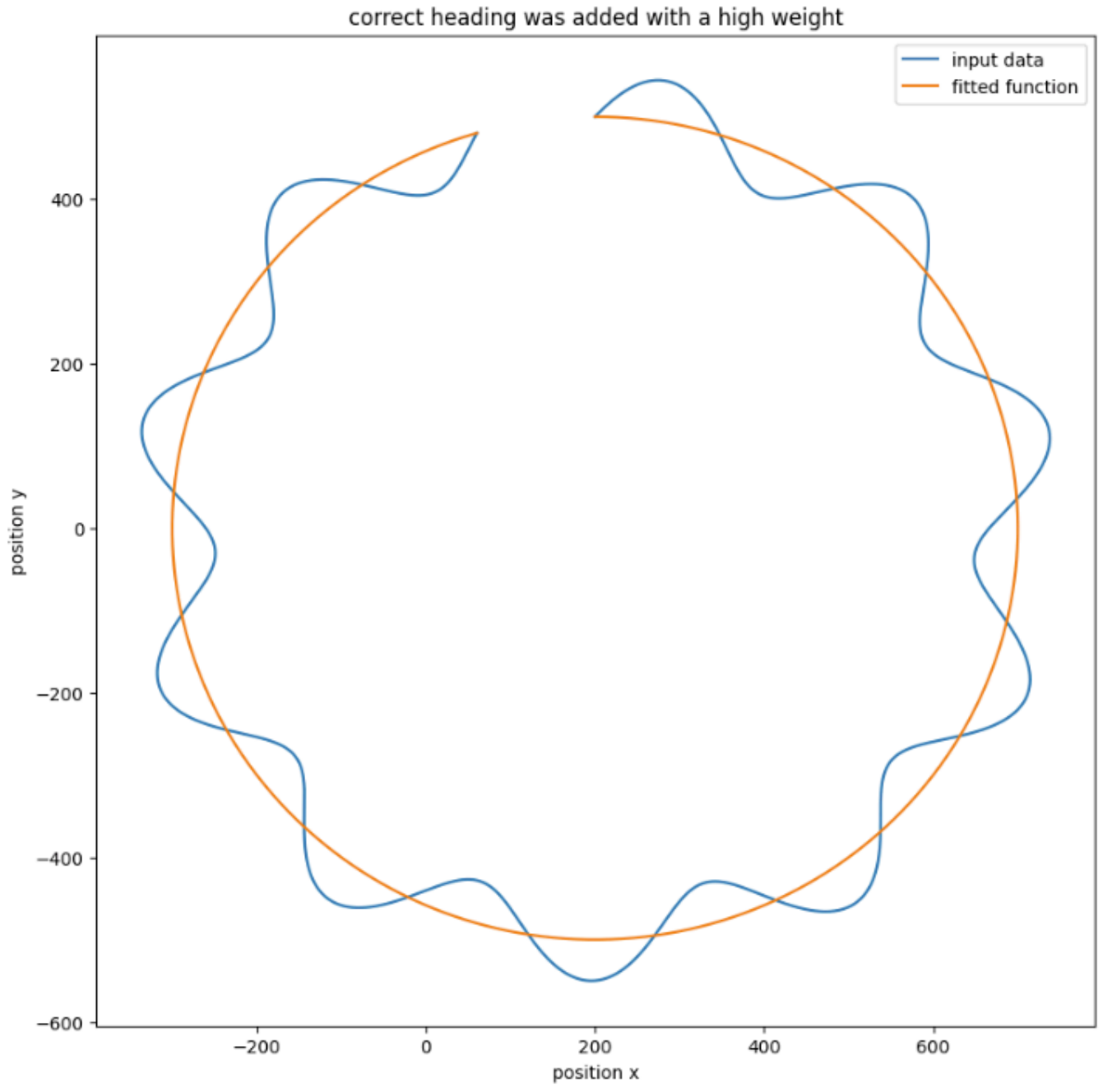

Fig. 4. *For this plot we are trying to fit a circle. However the measurment data of the position x and y was corrupted with a sinus wave. The sinus was chosen instead of high frequent white noise to mimic a different kind of error. By adding the heading from the normal circle with a heigh weight the fit follows the orgininal circle despite getting the position measurements from the corrupted circle.*

*D. Apriori Costs (Regularisation)*

Additionally, to measurement data we consider properties of the model Parameters itself or relations between different model parameters. This is typically referred to as regularization or filtering depending on context. The regularization should be chosen in a way to prefer even assumptions about the tracks and ensures a solvable problem for missing measurement data. Intuitively, the regularization incentivizes a simpler track is preferred even if a complex track would fit the measurements slightly better. Praxis shows that this is even more important if we want to estimate all track states (position, velocity, acceleration) only based on positions without directly measuring velocities and accelerations. Without any regularization the solution starts to swing, especially in the derivatives. General rule of thumb: the less measurements we have and the worse the quality of the measurement data the higher the importance of the regularization. For typical vehicle trajectories we can make the following assumptions.

- Smaller accelerations are more plausible
- Smaller acceleration changes are more plausible
- Locally smaller changes of changes are more plausible (curve drive)
- …

To easily formulate these conditions, we can exploit the structure of the physical spline. A smaller acceleration means that all coefficients except for the first two (initial position and initial velocity) are small. We call this a regularization of order zero of the physical spline. This results in the following

$$e_j = \frac{1}{2}\sum c_k \cdot w_k^2 \tag{18}$$

$$q_k = (0 \;\; \ldots \;\; 0 \;\; 1 \;\; 0 \;\; \ldots \;\; 0) \text{ with one at kth row}$$

$$Q = \sum c_k \cdot q_k \cdot q_k^T$$

$$or\ Q_j = c \cdot \begin{pmatrix} 0 & 0 & 0 & & & \\ 0 & 0 & & & 0 & \\ & 0 & 1 & & & \\ & & & & 1 & 0 \\ & 0 & & & 0 & 1 \end{pmatrix} \text{ for } c_k = c \text{ for all k.}$$

Depending on vehicle geometries, lateral and longitudinal acceleration limits are roughly on a similar scale. Thus a good and simple option is to weigh penalties on x and y direction with a similar weight.

Smaller acceleration changes would be called the order one regularization of the physical spline and be expressed with differences in the coefficients. Equation (19) adds costs for the difference between proceeding coefficients.

$$e_j = \frac{1}{2}\sum c_k \cdot (w_{k+1} - w_k)^2 \tag{19}$$

$$q_k = (0 \;\; \ldots \;\; 0 \;\; 1 \;\; -1 \;\; 0 \;\; \ldots \;\; 0) \text{ with one at kth row followed by} - 1$$

$$Q_j = \sum c_k \cdot q_k \cdot q_k^T$$

$$or\ Q_j = c \cdot \begin{pmatrix} 0 & 0 & & & & \cdot & & & \\ 0 & 0 & & & & \cdot & & & \\ & & 1 & -1 & 0 & \cdots & & \cdots & \\ & & -1 & 2 & -1 & 0 & \cdots & \cdots & \\ & & 0 & -1 & 2 & -1 & 0 & \cdots & \\ & & \cdot & 0 & -1 & 2 & -1 & 0 & \cdots \\ \cdots & & \cdots & 0 & \ddots & \ddots & \ddots & 0 & \cdots \\ & & \cdots & & 0 & \ddots & \ddots & \ddots & 0 \\ & & \cdots & & \cdot & 0 & -1 & 2 & -1 & 0 \\ & & \cdots & & \cdot & \cdot & 0 & -1 & 2 & -1 \\ & & \cdots & & \cdot & \cdot & \cdot & 0 & 1 & -1 \end{pmatrix}$$

$$for\ c_k = c\ for\ all\ k$$

Following the same scheme the order two regularization is given by

$$e_j = \frac{1}{2}\sum c_k \cdot \left((w_{k+2} - w_{k+1}) - (w_{k+1} - w_k)\right)^2 \tag{20}$$

$$Q_j = c \cdot \begin{pmatrix} 0 & 0 & & & & \cdot & & & & \\ 0 & 0 & & & & \cdot & & & & \\ & & 1 & -2 & 1 & 0 & \cdot & \cdot & \cdots & \\ & & -2 & 5 & -4 & 1 & 0 & \cdot & \cdots & \\ & & 1 & -4 & 6 & -4 & 1 & 0 & \cdots & \\ & & 0 & 1 & -4 & 6 & -4 & 1 & 0 & \cdot \\ \cdot & \cdot & \cdot & 0 & \ddots & \ddots & \ddots & \ddots & \ddots & 0 \\ & & \cdot & \cdot & 0 & 1 & -4 & 6 & -4 & 1 & 0 \\ & & \cdots & & & 0 & 1 & -4 & 6 & -4 & 1 \\ & & \cdots & & & \cdot & 0 & -2 & 5 & -4 & 1 \\ & & \cdots & & & \cdot & \cdot & 0 & 1 & -2 & 1 \end{pmatrix}$$

$$for\ c_k = c\ for\ all\ k.$$

In (Fig. 5) we can see the effect of a higher regularisation weight of order zero and one. Looking closely at the right end we can see that the slope of the velocity goes to zero for order zero and stays for order one. For order two the curvature would stay.

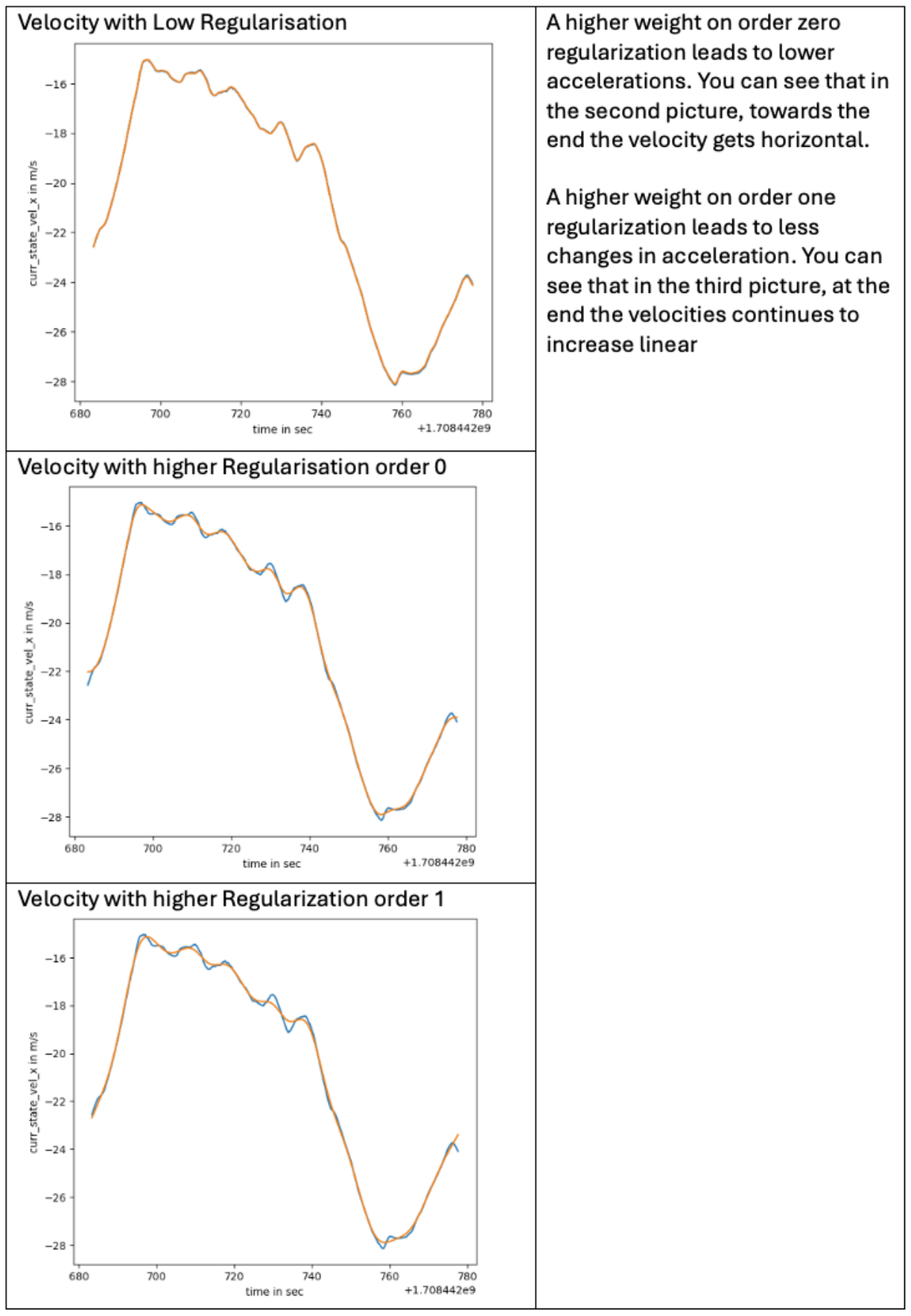

Fig. 5. *Effect of different regularisations*

Fitting the physical spline to standing objects could lead to a small shaking around zero velocity. It is mostly a cosmetic problem, because the effect is really small and has no cumulative drift. However lightly shaking standing objects are visually unpleasant when you generate a birds eye video of a moving car. To avoid that we introduce a regularization of order -1 to prefer smaller velocities. This regularisation is against the principle of momentum conservation. Thus this regularization will only be active during phases of longer intervals with zero velocity and skipped otherwise. The error function would be (21)

$$\begin{gathered} v_1 = w_1 \\ v_2 = w_1 + \Delta t \cdot w_2 \\ v_3 = w_1 + \Delta t \cdot (w_2 + w_3) \\ \dots \end{gathered} \tag{21}$$

$$e_j = \frac{1}{2} \cdot c_1 \cdot v_1^2 + \dots + \frac{1}{2} \cdot c_n \cdot v_n^2$$

$$\nabla e_j = c_1 \cdot \nabla v_1 \cdot v_1 + c_2 \cdot \nabla v_2 \cdot v_2 + \cdots$$

$$\nabla v_2 \cdot v_2 = c_2 \begin{pmatrix} 1 \\ \Delta t \\ \vdots \end{pmatrix} \cdot (w_1 + \Delta t \cdot w_2) = c_2 \cdot q_2^T q_2 \cdot w$$

$$\begin{gathered} q_1 = (1 \quad 0 \quad \dots \quad 0) \\ q_2 = (1 \quad \Delta t \quad 0 \quad \dots \quad 0) \\ q_3 = (1 \quad \Delta t \quad \Delta t \quad 0 \quad \dots \quad 0) \\ \dots \end{gathered}$$

$$Q_j = \sum c_k \cdot q_k^T \cdot q_k.$$

## V. Extensions for Heading

We can estimate the heading with the same model by introducing static nonlinearities, that get solved before and after the fitting. The functions $f_x$ and $f_y$ will be replaced with $f_{sin}$ and $f_{cos}$ . With this modelling we resolve the periodic nature of the heading. After the fitting the heading can be recreated with the following formular

$$f_{cos}(t) = \sum_j^N f_j(t) \cdot w_j \tag{22}$$

$$f_{sin}(t) = \sum_j^N f_j(t) \cdot w_j$$

$$\psi = \text{arctan2}\big(f_{cos}(t), f_{sin}(t)\big).$$

### A. Adding Heading measurements to the heading estimation

Equation (23) shows how to add heading measurements to the heading estimation.

$$e = \frac{1}{2}\sum \big(f_{\cos}(t_j) - \cos(\psi_j)\big)^2 + \frac{1}{2}\sum \big(f_{\sin}(t_j) - \sin(\psi_j)\big)^2 \tag{23}$$

Using the same procedure as above we get the following matrices

$$q_i = \begin{pmatrix} f_{sin,1}(t_j) \\ \vdots \\ f_{sin,n}(t_j) \\ f_{cos,1}(t_j) \\ \vdots \\ f_{cos,n}(t_j) \end{pmatrix} \quad Q_i = c_i \cdot q_i \cdot q_i^T \quad b_i = c_i \cdot \begin{pmatrix} f_{sin,1}(t_j) \cdot \sin(\psi_j) \\ \vdots \\ f_{sin,n}(t_j) \cdot \sin(\psi_j) \\ f_{cos,1}(t_j) \cdot \cos(\psi_j) \\ \vdots \\ f_{cos,n}(t_j) \cdot \cos(\psi_j) \end{pmatrix}. \tag{24}$$

### B. Adding velocity measurements to the heading estimation

Equation (25) shows how to add velocity measurements to the heading estimation

$$\begin{gathered} \dot{x} \cdot \sin(\psi) - \dot{y} \cdot \cos(\psi) = 0 \\ e = \frac{1}{2}\sum \left(\dot{x}_j \cdot f_{sin}(t_j) - \dot{y}_j \cdot f_{cos}(t_j)\right)^2. \end{gathered} \tag{25}$$

### C. Adding acceleration measurements to the heading estimation

Equation (26) shows how to consider acceleration measurements at a given velocity.

$$\begin{gathered} \dot{x} = v \cdot \cos(\psi) \quad \dot{y} = v \cdot \sin(\psi) \\ v = \sqrt{\dot{x}^2 + \dot{y}^2} \quad \dot{v} = \frac{\dot{x} \cdot \ddot{x} + \dot{y} \cdot \ddot{y}}{v} \\ \ddot{x} = \frac{d}{dt}(\cos(\psi)) \cdot v + \cos(\psi) \cdot \dot{v} \\ \ddot{y} = \frac{d}{dt}(\sin(\psi)) \cdot v + \sin(\psi) \cdot \dot{v} \\ e = \frac{1}{2}\sum \big(\dot{f}_{cos}(t_j) \cdot v_j + f_{\cos}(t_j) \cdot \dot{v}_j - \ddot{x}_j\big)^2 \\ + \frac{1}{2}\sum \big(\dot{f}_{sin}(t_j) \cdot v_j + f_{\sin}(t_j) \cdot \dot{v}_j - \ddot{y}_j\big)^2 \end{gathered} \tag{26}$$

## VI. Results and Use Cases

### A. Post processed trajectory optimization

Our main use case for the physical Spline is the improvement of recorded object tracks in the post processing. The tracks were recorded with the sensors of an autonomous vehicle (Lidar, Radar, Camera). The results can later be used for analysis or as pseudo ground truth for machine learning models. The physical spline delivers kinematically consistent object tracks that make use of the advantage that in post processing we already have access to the whole object track. In case of an online implementation the physical spline would have to be implemented as a moving horizon estimator (see further in [5]). To tune the physical spline, we equipped another vehicle with a high precision GNSS system that was optimized later with a PPK approach. The recorded data was used as benchmark to compare the recorded Lidar/Radar/Camera measurements. The results look very promising. Fig 6 shows the raw data in blue and the improved data with the physical spline in orange. Each subplot shows another signal of the object track. In Fig. 7 we can see additionally in green the ground truth of the high precision measurement device that was mounted inside of the tracked vehicle for reference. We can see that the physical spline corrects many errors. Due to its consistency with the physics of the vehicle, the jumps in the input measurement signals were not possible and dominated by the correct influence of other measured signals. The worst performance we can see at the corner of the object track. Especially the heading drops of towards the end. The reason for that is that the measurements are no longer surrounded and backed up by additional data from both sides.

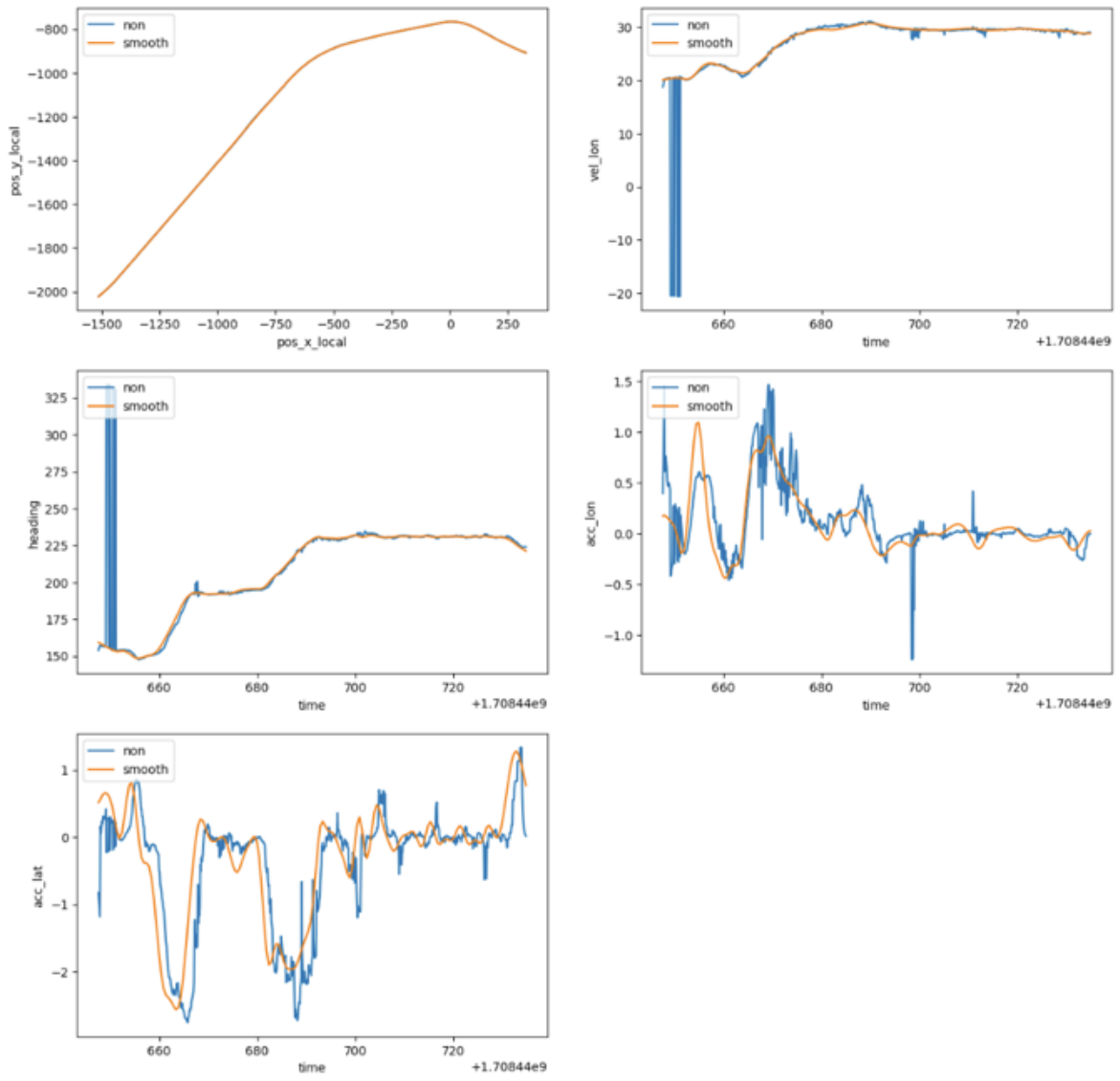

Fig. 6. *Smoothed state variables and raw recorded trajectorie. Top left x-y brids eye plot, top right longitudinal velocity, middle left heading, middle right longitudinal acceleration, bottom left lateral acceleration.*

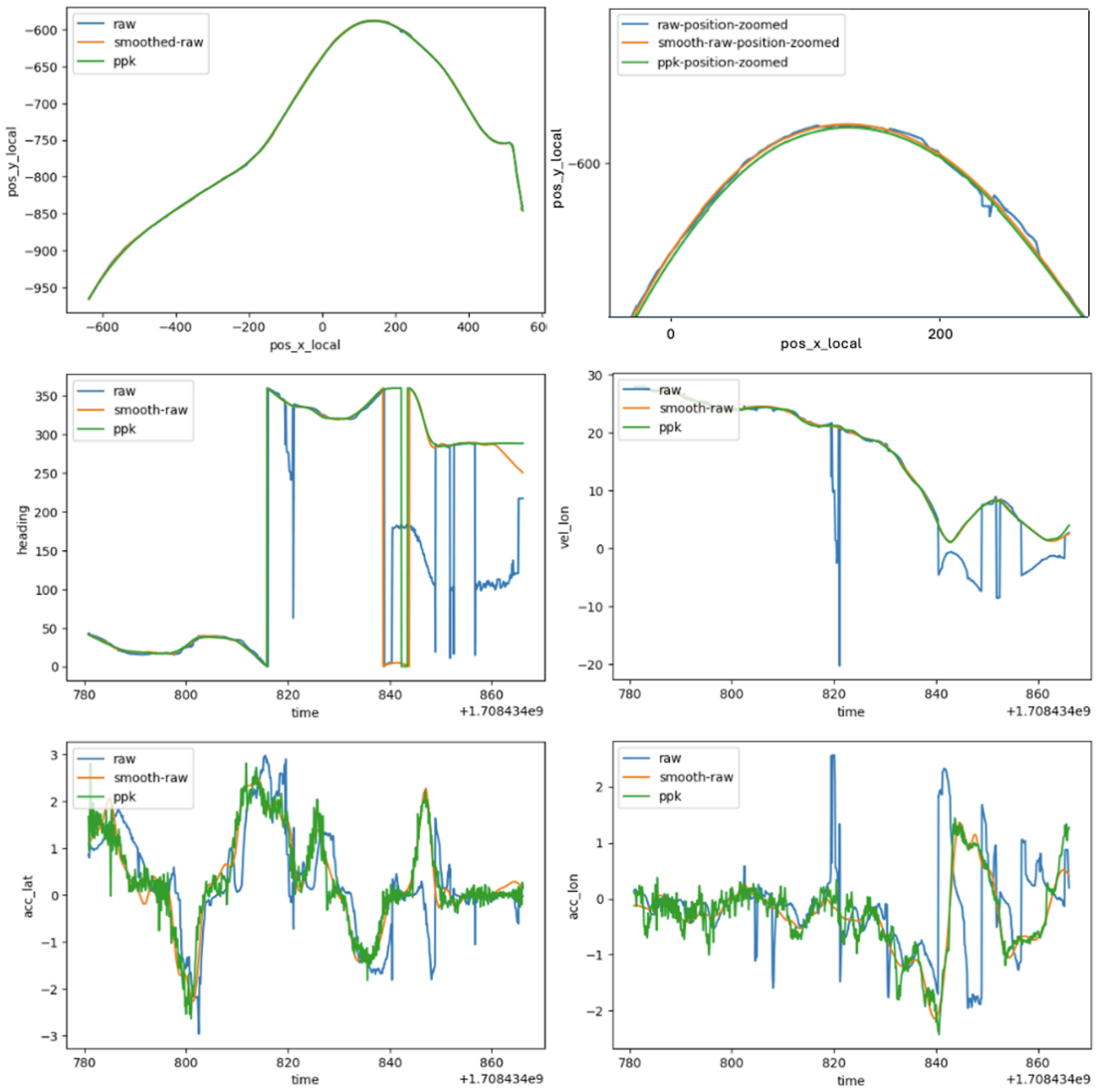

Fig. 7. *raw, physical spline optimized and ground truth ppk object track. Top left and right shows the x-y birds eye plot, middle right the longitudinal velocity, middle left the heading, bottom right the longitudinal acceleration and bottom left the lateral acceleration*

## B. *Upsampling and object track reconstruction*

For Certain data sources we only have position measurements. We can run the physical spline with positions only and the regularization parameters to sample on another time vector and create velocities and accelerations.

## C. *Data Compression*

With the physical spline the whole trajectory is described by the parameter vector. By only storing the parameter vector we can recreate the hole trajectory.

# VII. NUMERICAL NOTES/CHALLENGES

Most of the computational costs occur for building up the matrix and evaluating the basis functions. Solving the lgs itself is fast in comparison.

Typically, we want to add many measurements at once to the optimization Problem, thus we should vectorize the evaluation of the basis function, such that a complete time vector can be handed at once. The return should be a Matrix. One dimension is for different timestamps and the other for the different basis functions (or derivative or other formulations from the dyadic products above)

$$\begin{pmatrix} f_0(t_0) & \dots & f_N(t_0) \\ \vdots & \ddots & \vdots \\ f_0(t_M) & \dots & f_N(t_M) \end{pmatrix} \qquad Q_j = \sum c_k \cdot q_k^T \cdot q_k \qquad q_k = \big(f_0(t_k) \dots f_N(t_k)\big) = kth - row. \tag{27}$$

To build up the Matrix we must sum up the dyadic products of the rows. To do this more efficiently we can use the following theorem (28)

$$A \cdot B = \sum_{k=1}^{N} c_{A,k} \cdot r_{B,k} \tag{28}$$

Where $c_{A,k}$ are the columns of $A$ and $r_{B,k}$ are the rows of $B$, while $N$ denotes the number of columns of $A$ (the same as the number of rows of $B$, such that the multiplication could be performed) [6]. Now we can just multiply the matrix from (27) with its transposed.

# VIII. CONCLUSION AND EXTENSIONS

The achieved results can be scaled and achieve good outcomes. For possible extensions we could add equality and inequality constraints to the quadratic optimization to enforce a specific behavior.

We could also use a nonlinear more complex model that keeps the idea of the physical meaning of the basis functions. We could also use a model that estimates the heading at the same time. The more complex models would require working with iterative solvers like gradient descent.